\documentclass[aps, nofootinbib,twocolumn]{revtex4}
\usepackage{amsmath}%,amssymb}

\usepackage{amssymb}
\allowdisplaybreaks

\usepackage{makeidx}
\usepackage{amsfonts}
\usepackage[usenames,dvipsnames]{pstricks}
\usepackage{subfigure}
\usepackage{epsfig}
\usepackage{pst-grad} % For gradients
\usepackage{mathtools}
\usepackage[colorlinks,hyperindex]{hyperref}
\usepackage{array}

\hypersetup
{	colorlinks,%
	citecolor=black,%
	linkcolor=black,%
	urlcolor=black,%
}

\usepackage{allpurpose}

\usepackage{flushend}% Align the end of the last page.

\newcommand\nn{{\nonumber}}

\begin{document}

\title{Perturbative deflection angle, gravitational lensing in the strong field limit and the black hole shadow}

\author{Junji Jia}
\email{junjijia@whu.edu.cn}
\affiliation{MOE Key Laboratory of Artificial Micro- and Nano-structures, School of Physics and Technology, Wuhan University, Wuhan, 430072, China}

\author{Ke Huang}
\affiliation{MOE Key Laboratory of Artificial Micro- and Nano-structures, School of Physics and Technology, Wuhan University, Wuhan, 430072, China}

%\address[2]{Center for Astrophysics \& MOE Key Laboratory of Artificial Micro- and Nano-structures, School of Physics and Technology, Wuhan University, Wuhan, 430072, China}

\date{\today}

\begin{abstract}
A perturbative method to compute the deflection angle of both timelike and null rays in arbitrary static and spherically symmetric spacetimes in the strong field limit is proposed. The result takes a quasi-series form of $(1-b_c/b)$ where $b$ is the impact parameter and $b_c$ is its critical value, with coefficients of the series explicitly given. This result also naturally takes into account the finite distance effect of both the source and detector, and allows to solve the apparent angles of the relativistic images in a more precise way. From this, the BH angular shadow size is expressed as a simple formula containing metric functions and particle/photon sphere radius. The magnification of the relativistic images were shown to diverge at different values of the source-detector angular coordinate difference, depending on the relation between the source and detector distance from the lens. To verify all these results, we then applied them to the Hayward BH spacetime, concentrating on the effects of its charge parameter $l$ and the asymptotic velocity $v$  of the signal. The BH shadow size were found to decrease slightly as $l$ increase to its critical value, and increase as $v$ decreases from light speed. For the deflection angle and the magnification of the images however, both the increase of $l$ and decrease of $v$ will increase their values. 
\end{abstract}


\maketitle

\section{Introduction}

Geodesic behavior of both null and timelike rays in the strong field limit (SFL) near the ultra-compact objects, such as black holes (BHs) and neutron stars (NS), is important for many phenomena. The accretion of materials \cite{Bambi:2012tg}, the shape of BH shadow \cite{Grenzebach:2014fha}, the supernova explosion details \cite{Ott:2012mr} and the ring-down phase of BH/NS binary mergers \cite{Assumpcao:2018bka} are among the many such phenomena.
In order to understand these well, to test BH and NS physics and to test GR in the SFL, therefore it is necessary to understand the general geodesic behavior of null/timelike rays near the gravitational center.

Previously, the null geodesics in this limit has been well studied and understood in many spacetimes. The existence of photon sphere (surface) around some BHs \cite{Claudel:2000yi} has been well known, and the location of the innermost stable circular orbits \cite{Psaltis:2008bb, Abramowicz:2011xu} has been intensively investigated.
In particular, the deflection of null rays in the SFL has been systematically studied by Bozza et al. for static and spherically symmetric (SSS) spacetimes and equatorial motion in stationary and axisymmetric (SAS) spacetimes \cite{Bozza:2002zj,Bozza:2007gt,Bozza:2009yw} and has since been used to study the deflection and GL in the SFL for many interesting spacetimes \cite{Perlick:2004tq,Whisker:2004gq, Bozza:2005tg,Eiroa:2005ag,Nandi:2006ds, Chen:2009eu,Stefanov:2010xz,Tsukamoto:2012xs, Sahu:2012er, Wei:2014dka}. Note that most of the deflection angles found in these works assumed that the source and detectors are located at infinite distance. Through these studies, it is found that when the impact parameter $b$ approaches the critical value $b_c$, the deflection angle diverges as
\be
\alpha(b\to b_c^+)\approx \bar{a}\log \lb 1-\frac{b}{b_c}\rb +\bar{b} +~\text{higher order terms},
\label{eq:defanginsfl}
\ee
where $\bar{a},~\bar{b}$ are some coefficients. With this deflection angle, then the single most remarkable consequence is the existence of a series of relativistic image with infinite magnification for rays near the critical photon sphere.

The messenger for these GL images and for GL in the weak field limit were traditionally light rays. However, with the discovery of extragalactic neutrinos from SN1987A \cite{Hirata:1987hu, Bionta:1987qt} and blazar TXS 0506+056 \cite{IceCube:2018dnn,IceCube:2018cha} and the gravitational waves (GWs) \cite{Abbott:2016blz,Abbott:2016nmj,Abbott:2017oio,TheLIGOScientific:2017qsa,Monitor:2017mdv}, and especially the lensed supernovas \cite{Kelly:2014mwa,Goobar:2016uuf} and simultaneous observation of GW+GRB events \cite{TheLIGOScientific:2017qsa,Monitor:2017mdv}, we know that timelike rays might also be messengers for GL or other gravitational effects in the SFL. Therefore in this work we will investigate the influence of the asymptotic velocity on the deflection angle and GL in the SFL for general SSS spacetimes.
Moreover, we adapt a perturbative method developed previously for the calculation of the deflection angle and time delay in the weak field limit \cite{Huang:2020trl, Liu:2020wcu, Liu:2020mkf,Duan:2020tsq, Jia:2020xbc} to the SFL case, and use it to find a series form of the deflection angle for source and detector at finite distances. The leading two orders of this series will produce the deflection angle \eqref{eq:defanginsfl} when the asymptotic velocity is set to light speed and source/detector distances set to infinity. Therefore results in this work will be more general in that it includes the effect of timelike velocity, the finite distance of the source and detector and contributions from orders higher than constant. 

The paper is organized as the following. In Sec. \ref{sec:pertmeth} we describe in detail this perturbative method. The result of the deflection angle is given in Eq. \eqref{eq:alphares}-\eqref{eq:alphacdef} and its realizations in some simple spacetimes are compared with known results. In Sec. \ref{sec:glbhsz}, we study the GL, including the apparent angles and the magnification, and BH shadow size  in the SFL for signals with arbitrary velocity and source/detector from finite distances. In Sec. \ref{sec:rn}, the perturbative method and  the GL results then are applied to the Hayward spacetime  and the effects of the charge parameter and signal velocity are analyzed. We summarize and discuss possible extensions in the end.

\section{Perturbative method\label{sec:pertmeth}}

We start from the most general SSS metric, which is described by
\be
\dd s^2=-A(r)\dd t^2+B(r)\dd r^2+C(r)\lb \dd {\theta}^2+\sin^2{\theta} \dd \varphi^2\rb \label{sphmetric}
\ee
where $(t,~r,~{\theta},~\varphi)$ are the coordinates and $A,~B,~C$ are metric functions depending on $r$ only. The function $C(r)$ can be further chosen to $r^2$ (at least locally). However, sometimes in special coordinate systems this is not explicitly set so we will keep it free here.

The corresponding geodesic equations are
\begin{subequations}\label{eq:geoeqs}
\begin{align}
    \dot{r}^2&=\frac{\lb \frac{E^2}{A(r)}-\kappa\rb C(r)-L^2}{C(r)B(r)}, \label{eq:dotreq}\\
    \dot{\phi}&=\frac{L}{C(r)},\\
    \dot{t}&=\frac{E}{A(r)},
\end{align}
\end{subequations}
where without losing any generality we have set ${\theta}=\pi/2$. $\kappa=0,~1$ for massless and massive particles respectively.
Here $L$ and $E$ are the angular momentum and energy of the signal (per unit mass) respectively. They can be related to the impact parameter $b$ of the signal and its velocity $v$ at infinity by
\be
L=\frac{b}{v}{\sqrt{1-v^2}},~E=\frac{1}{\sqrt{1-v^2}}. \label{eq:ledef}
\ee

The change of the angular coordinate from source located at radius $r_s$ to detector at radius $r_d$ then is
\bea
\Delta \phi&=&\sum_{i=s,d}\int_{r_0}^{r_i}\frac{\dot{\phi}}{\dot{r}}\dd r\nn\\
&=&\sum_{i=s,d}\int_{r_0}^{r_i}\sqrt{\frac{B(r)}{C(r)}}\frac{L \dd r}{\sqrt{\lb \frac{E^2}{A(r)}-\kappa\rb C(r)-L^2}}.\label{eq:deflection}
\eea
We assume that the detector is located far away from the gravitational center while $r_s$ is not necessarily so. Here $r_0$ is the closest approach defined by $\dot{r}|_{r=r_0}=0$ which after using Eq. \eqref{eq:dotreq} becomes
\be
\lb \frac{E^2}{A(r_0)}-\kappa\rb C(r_0)-L^2=0. \label{eq:tr0ldef}
\ee
This $r_0$ can also be related to the impact parameter $b$ by using Eq. \eqref{eq:ledef} and \eqref{eq:tr0ldef}
\be\label{eq:binr0}
\frac{1}{b}=\frac{\sqrt{E^2-\kappa}}{L}=\sqrt{\frac{E^2-\kappa}{\lb \frac{E^2}{A(r_0)}-\kappa\rb C(r_0)}}.
\ee

We also assume that the spacetime contains a particle (or photon) sphere below which the particle (or photon) will spiral into the BH and not escape. This particle sphere has a radius $r_c$ defined as the critical point of the denominator of Eq. \eqref{eq:deflection}
\begin{equation}
    \dd \lsb\lb \frac{E^2}{A(r)}-\kappa\rb C(r)\rsb/\dd r\Bigg|_{r=r_c}=0. \label{eq:rcdef}
\end{equation}
If the closest approach $r_0$ equals particle sphere radius $r_c$, then we call the impact parameter corresponding to this $r_0$ the critical impact parameter $b_c$. That is, using Eq. \eqref{eq:binr0},
\be
\frac{1}{b_c}=\sqrt{\frac{E^2-\kappa}{\lb \frac{E^2}{A(r_c)}-\kappa\rb C(r_c)}}. \label{eq:bcdef}
\ee
For example, for Schwarzschild spacetime, it is well known that for light signal with $\kappa=0$, the photon sphere has a radius $r_c=3M$ as predicted by Eq. \eqref{eq:rcdef} and a critical impact parameter $b_c=3\sqrt{3}M$ which can be verified using Eq. \eqref{eq:bcdef}. While for light signal in RN BH spacetime, it is known that \cite{Pang:2018jpm}
\bea
  r_c&=&\frac{(3+\sqrt{9-8\hat{Q}^2})M}{2},~\text{where}~\hat{Q}\equiv Q/M,\label{rchv1}\\
  b_c&=&M\sqrt{\frac{4 \hat{Q}^2(2\hat{Q}^2-9)+\left(9-8 \hat{Q}^2\right)^{3/2}+27}{2(1- \hat{Q}^2)}}.\label{eq:bchv1}
\eea

We now define a function $p(x)$ that is inspired by Eq. \eqref{eq:bcdef}
\be
p\lb \frac{1}{r_0}\rb =\frac{1}{b_c}-\sqrt{\frac{E^2-\kappa}{\lb \frac{E^2}{A(r_0)}-\kappa\rb C(r_0)}} \label{eq:p1or0}
\ee
and denote its inverse function as $q(x)$. Clearly, these functions satisfy $ p \lb \frac{1}{r_0}\rb =\frac{1}{b_c}-\frac{1}{b}$
and
\be \frac{1}{r_0}  =q\lb \frac{1}{b_c}-\frac{1}{b}\rb =q\lb \frac{ 1-b_c/b}{b_c}\rb  .\label{eq:pqprop2}\ee
Using function $q(x)$,  for $\Delta \phi$ in Eq. \eqref{eq:deflection} we then do a change of integration variables from $r$ to $\xi$ linked by
\be \label{eq:chavdef}
\frac{1}{r}=q\lb \frac{\xi}{b_c}\rb ,~\mbox{i.e.,}~p\lb \frac{1}{r}\rb =\frac{\xi}{b_c}\ee
such that after using Eqs. \eqref{eq:ledef}, \eqref{eq:p1or0} and \eqref{eq:pqprop2} the various terms in the integral \eqref{eq:deflection}  becomes respectively,
\begin{subequations}\label{eq:changesunderchanges}
\begin{align}
&r_0\to 1-b_c/b,\\
&r_{s,d}\to 1-b_c\sqrt{\frac{E^2-\kappa}{\lb \frac{E^2}{A(r_{s,d})}-\kappa\rb C(r_{s,d})}}\equiv \eta_{s,d},\label{eq:etedef}\\
&\lb \frac{E^2}{A(r)}-\kappa\rb C(r)\to \frac{b_c^2(E^2-\kappa)}{(\xi-1)^2},\\
&L\to b\sqrt{E^2-\kappa},\\
&\sqrt{\frac{B(r)}{C(r)}}\to \sqrt{\frac{B(1/q)}{C(1/q)}},\\
&\dd r\to -\frac{q^\prime}{q^2}\dd \xi,
\end{align}
\end{subequations}
where $q=q(\xi/b_c)$ and $q^\prime$ is its derivative. For quantity $\eta_{s,d}$, it is through them that $\Delta\phi$ depends on the finite distance $r_s$ and $r_d$. Also note that $\arcsin(1-\eta_{d})$ is the apparent angle of the GL images observed by the detector (see Ref. \cite{ourletter,bk:gr2}) and in the infinite $r_{s,d}$ limit, we will have $\eta_{s,d}=1$ because $C(r_{s,d}\to\infty)\to r_{s,d}^2$. The change of variable \eqref{eq:changesunderchanges}  is the key step that enables the later integrability of $\Delta\phi$ in subsection \ref{subsec:appone}.
Note that the dependence of $\Delta\phi$ on the finite distance $r_s$ and $r_d$ are through the $\eta_{s,d}$.  Organizing the terms in Eq. \eqref{eq:changesunderchanges} together, Eq. \eqref{eq:deflection} is transformed to
\begin{equation}\label{deflectionlambda}
    \Delta\phi=\sum_{i=s,d}\int_a^{\eta_i}\sqrt{\frac{B(1/q)}{C(1/q)}} \frac{\xi-1}{b_c\sqrt{(2-a-\xi)(\xi-a)}}
         \frac{q^\prime}{q^2}\dd \xi
\end{equation}
where $a\equiv 1-b_c/b$.
Then our target for calculating $\Delta \phi$ in the SFL becomes to calculate this integral in the limit $b\to b_c^+$, i.e., $a\to 0^+$.

\subsection{Perturbative Computation of $\Delta \phi$ \label{subsec:appone}}

To proceed from Eq. \eqref{deflectionlambda}, we will denote the integrand of Eq. \eqref{deflectionlambda}, except the factor $1/\sqrt{\xi-a }$, as
\be y(\xi,a) = \sqrt{\frac{B(1/q )}{C(1/q )}}  \frac{\xi-1}{b_c\sqrt{2-a-\xi}} \frac{q^\prime}{q^2}.
\label{eq:ydef}\ee
Since we are interested in the small $a$ limit of the integral, to which the contribution from $y(\xi,a)$ when $\xi$ is small is expected to be large, we can directly expand $y(\xi,a)$ for small $\xi$ and then attempt to carry out the integral.
The square-root factor of $y(\xi,a)$ has the following expansion
\be
\frac{1}{\sqrt{(2-a-\xi)}}= \sum_{n=0}^\infty \frac{(2n-1)!!}{(2n)!!}\frac{\xi^n}{(2-a)^{n+\frac12}}.\label{eq:exp1f}
\ee
Denoting the rest factors of $y(\xi)$ as $f(\xi)$,
\be
f(\xi)=\sqrt{\frac{B(1/q )}{C(1/q )}}  \frac{\xi-1}{b_c} \frac{q^\prime}{q^2}
,\ee
then one can show that it has an expansion
\be f(\xi)=\sum_{n=-1}^\infty f_n \xi^{\frac{n}{2}},\ee
where the initial index is -1 due to the $q^\prime$ factor in $y(\xi)$ and $f_n$ are the coefficients. We can then combine the above two expansions and collect according to the power of $\xi^n$
\be
y(\xi)=\sum_{n=-1}^\infty  \lsb \sum_{m=0}^{\lsb \frac{n+1}{2}\rsb} \frac{ a^m}{(2-a)^{\lsb \frac{n+1}{2}\rsb +\frac12}} y_{n,m} \rsb \xi^{n/2} \label{eq:yexpded}
\ee 
where $y_{n,m}$ denote coefficients of corresponding powers of $\xi, ~a$ and $(2-a)$.
Note that these $y_{n,m}$ are neither $\xi$ nor $a$ dependant and can be directly worked out once the metric functions are known, as will be shown in subsection \ref{subsec:appone}. The $\Delta \phi$ in Eq. \eqref{deflectionlambda} consequently becomes
\be
\Delta\phi=\sum_{i=s,d}\sum_{n=-1}^\infty\sum_{m=0}^{\lsb \frac{n+1}{2}\rsb} \frac{a^m}{(2-a)^{\lsb\frac{n+1}{2}\rsb +\frac12}} y_{n,m}\int_a^{\eta_i}  \frac{\xi^{n/2}}{\sqrt{\xi-a}}\dd \xi. \label{eq:defang4}
\ee

At this point, the integrability of $\Delta\phi$ becomes apparent because the integrals of the form
$\displaystyle  \int_a^{\eta_i}  \frac{\xi^{n/2}}{\sqrt{\xi-a}}\dd \xi~(n=-1,0,cdots)$ can always be carried out. For even and odd $n$ respectively, the results of such integrals are given by the Eq. \eqref{eq:ttgenintmk} in Appendix \ref{app:ellp}.
Substituting them into Eq. \eqref{eq:defang4}, $\Delta \phi$ finally becomes
\begin{widetext}
\bea
\Delta\phi&=&\sum_{i=s,d}\sum_{k=0}^\infty \sum_{m=0}^k \frac{ a^m}{(2-a)^{ k+\frac12}}
\lcb y_{2k-1,m} \cdot\frac{a^kC_{2k}^k}{4^k}\lsb -\ln a +2\ln\lb \sqrt{ \eta_i}+\sqrt{\eta_i-a}\rb + \sum_{j=1}^{k}\frac{4^j}{jC_{2j}^{j}}\lb \frac{\eta_i}{a}\rb^j\sqrt{1-\frac{a}{\eta_i}}\rsb \right.\nn\\
&&\left. +y_{2k,m}\sum_{j=0}^{k} \frac{2 C_{j}^{k}a^{k-j}\lb \eta_i-a\rb^{j+1/2}}{2j+1}\rcb. \label{eq:alphares}
\eea
It is seen that the only divergence in the SFL is proportional to $\ln a$ and caused by the $k=m=0$ term.
In the small $a$ limit, all functions of $a$ here can be further expanded and then this $\Delta \phi$ can be thought as a quasi-series of $a$, with only one term in the coefficient of each $a^n~(n=0,~1,~\cdots)$  containing one $\ln a$. That is,
\be
\Delta\phi=\sum_{n=0}^\infty \lsb -C_n\ln a+D_n\rsb  a^n, \label{eq:dphgform}
\ee
where $-C_n$ and $D_n$ are coefficients depending on the spacetime behaviour near the particle sphere through $y_{k,m}$, on the finite distance $r_{s,d}$  through $\eta_{s,d}$ and on the particle velocity $v$ through both. To the $\mathcal{O}(a)^0$ order, only the $m=0,~j=k$ terms in Eq. \eqref{eq:alphares} contribute to $\Delta \phi$, i.e.,
\bea
\Delta\phi&=&\sum_{i=s,d} \lcb
\frac{\sqrt{2}y_{-1,0}}{2}\lsb -\ln a +2\ln\lb 2\sqrt{ \eta_i}\rb\rsb  +\sum_{n=0}^\infty \frac{2y_{n,0}  \eta_i^{\frac{n+1}{2}}} {2^{\lsb\frac{n+1}{2}\rsb+\frac12}(n+1)}\rcb +\mathcal{O}(a)^1 \label{eq:alpharesordera0}\\
&= & -C_0 \ln a+D_0(\eta_s,\eta_d) +\mathcal{O}(a)^1 \label{eq:alphacdef}
\eea
\end{widetext}
where $C_0$ and $D_0(\eta_s,\eta_d)$ can be read off directly from Eq.  \eqref{eq:alpharesordera0} to be
\bea
&&C_0=\sqrt{2}y_{-1,0},\label{eq:clapp1}\\
&&D_0=\sum_{i=s,d}\lsb \sqrt{2}\ln(2\sqrt{\eta_i})
+\sum_{n=0}^\infty \frac{2y_{n,0}  \eta_i^{\frac{n+1}{2}}} {2^{\lsb\frac{n+1}{2}\rsb+\frac12}(n+1)}\rsb. \label{eq:c0app1}
\eea

To link $y_{n,m}$ to the behavior of the spacetime in a more transparent way, we will first express them using expansions of the metric functions around the particle sphere. It will be seen that for any particular $n$, the $y_{n,m}$ are linked to the metric expansion coefficients up to the $n$-th order. And then in Table \ref{tb:comp}, we will apply these results to particular spacetimes and compare $\Delta \phi$ found using our method to known analytic formulas and numerical values.

Assuming that the metric functions $A(r),~B(r)$ and $C(r)$ have the following series expansion at the particle sphere $r=r_c$
\begin{subequations}\label{eq:metricrcexp}
\begin{align}
&A(r\to r_c)=\sum_{n=0}^\infty a_n(r-r_c)^n,\\
&B(r\to r_c)=\sum_{n=0}^\infty b_n(r-r_c)^n,\\
&C(r\to r_c)=\sum_{n=0}^\infty c_n(r-r_c)^n, \label{eq:crexp}
\end{align}
\end{subequations}
then we see that the very fact that $r_c$ is the radius of the particle sphere demands through Eq. \eqref{eq:rcdef} that
\be \label{eq:rccons}
c_1 \left(\frac{E^2}{a_0}-\kappa \right)-\frac{a_1 c_0 E^2}{a_0^2}=0.
\ee
This condition provides a constraint from which we can express $a_1$ in terms of $a_0,~c_0$ and $c_1$ 
\be 
a_1= \frac{a_0^2c_1}{c_0 E^2} \left(\frac{E^2}{a_0}-\kappa \right) \label{eq:a1inothers}
.\ee
Moreover, the critical impact parameter in Eq. \eqref{eq:bcdef} can also be related to these coefficients
\begin{equation}
\displaystyle
b_c=\sqrt{\frac{c_0(\frac{E^2}{a_0}-\kappa)}{E^2-\kappa}}
\end{equation}
It is also interesting to note that this value depends only on the metric functions $A(r)$ and $C(r)$ but not $B(r)$.
Furthermore, without losing any generality, $C(r)$ can be set to $r^2$. In this case, the expansion \eqref{eq:crexp} of $C(r)$ only contains three terms
\be
c_0=r_c^2,~c_1=2r_c,~c_2=1,~c_n=0~(n\geq 3). \label{eq:c0to2}\ee
If the metric function $B(r)=1/A(r)$ as in many SSS cases, then the entire set of $B(r)$'s coefficients $b_n~(n=0,~1,~\cdots)$ can be correlated with that of $A(r)$ and $r_c$. In these cases, the number of independent coefficients will be much lesser.

Using the expansions \eqref{eq:metricrcexp}, together with Eqs. \eqref{eq:p1or0}, \eqref{eq:ydef}, \eqref{eq:yexpded} and the Lagrangian inverse theorem, we can compute the coefficients $y_{n,m}$ to any desired order. The first few $y_{n,0}$ which are crucial for $\Delta\phi$ to the $\mathcal{O}(a)^0$ order are
\begin{subequations}\label{eq:y0nres}
\begin{align}
y_{-1,0}&=b_c \sqrt{\frac{b_0 }{2c_0 T_2}}, \label{eq:ym10nres}\\
y_{0,0}&=\frac{b_c^2 \left(b_1 c_0 T_2-b_0 c_1 T_2-2 b_0 c_0 T_3\right)}{ 2\sqrt{b_0} c_0^{3/2} T_2^2}
\end{align}
\end{subequations}
where constants $T_2$, $T_3$ are
\begin{subequations}\label{eq:tnres}
\begin{align}
%T_0=&\frac{c_0E^2}{E^2-\kappa}\left(\frac{1}{a_0}-\frac{\kappa}{E^2}\right),\\
T_2=&\frac{E^2}{E^2-\kappa}\lsb c_2\left(\frac{1}{a_0}-\frac{\kappa}{E^2}\right)+\frac{\left(a_1^2-a_0 a_2\right) c_0 }{a_0^3}-\frac{a_1 c_1 }{a_0^2}\rsb,\\
T_3=&\frac{E^2}{E^2-\kappa} \lsb c_3\left(\frac{1}{a_0}-\frac{\kappa}{E^2}\right)
+\frac{\left(a_1^2-a_0 a_2\right) c_1 }{a_0^3}\right.\nn\\
&\left.
-\frac{a_1 c_2 }{a_0^2}-\frac{\left(a_1^3-2 a_0 a_1a_2 +a_0^2 a_3\right) c_0 }{a_0^4}\rsb.
\end{align}
\end{subequations}
It is seen that $y_{-1,0}$, which determines the divergence as $a\to 0^+$, is fixed by the metric expansion coefficients $a_{0, 2},~b_0$ and $c_{0,1,2}$ but not the higher order coefficients. Higher order $y_{n,0}$ ($n= 1,2$) are given in Appendix \ref{app:ellp} in Eq. (\ref{eq:y0nres}c-d) because of their excessive length. From these, one finds that for a general $n$, $y_{n,0}$ are determined by coefficients up to $a_{n+3},~b_{n+1}$ and $c_{n+3}$. Once the spacetime metric is known and its $r_c$ radius is found, then all these $y_{n,0}$ can be readily obtained from Eq. \eqref{eq:y0nres}.

\subsection{Comparison with known results}

Now we can apply the above Eqs. \eqref{eq:alpharesordera0}-\eqref{eq:c0app1}  for $\Delta\phi$  to particular spacetimes to verify the validity of our methodology by comparing our results to others in the SFL. In Table \ref{tb:comp}, we list the spacetimes and the quantities we compared, the method for comparison (analytical or numerical) and the corresponding references. All these literatures only worked for null signal and most of the them only presented deflection angle numerically.  In all literature compared, our Eq. \eqref{eq:alpharesordera0} were able to reproduce the corresponding quantities in that spacetime. 

\begin{table*}
\begin{tabular}{l|l|l|l}
\hline\hline
Spacetime& Quantities & Method & Reference\\
\hline
Schwarzschild&$C_0,~D_0$&Analytical, Numerical&\cite{Iyer:2006cn}\\
RN&$C_0,~D_0$&Numerical&\cite{Bozza:2002zj,Whisker:2004gq,Chen:2009eu}\\
Braneworld BH&$C_0,~D_0$&Numerical&\cite{Whisker:2004gq,Majumdar:2005ba}\\
JNW&$C_0,~D_0,~\theta_{\infty,s} $&Numerical&\cite{Nandi:2006ds}\\
GMCFM \uppercase\expandafter{\romannumeral2}&$C_0,~D_0$&Numerical&\cite{Nandi:2006ds,Majumdar:2005ba}\\
GMGHS&$\theta_{\infty,s},~\theta_{1,s}$&Numerical&\cite{Bhadra:2003zs}\\
Bardeen&$C_0,~D_0$&Numerical&\cite{Eiroa:2010wm}\\
\hline
\hline
\end{tabular}
\caption{Comparison of our work and known literatures. The columns are respectively the spacetime name, the quantities compared, the method used for comparison and the corresponding reference. Note that the coefficient $D_0$ in Ref. \cite{Chen:2009eu}  (denoted as $\bar{b}$) seems at odds with all other works. JNW, GMCFM and GMGHS stands for the Janis–Newman–Winicour, Germani-Maartens-Casadio-Fabbri-Mazzacurati and Gibbons-Maeda-Garfifinkle-Horowitz-Strominger spacetimes respectively. See Eq. \eqref{eq:thetanres} for meaning of $\theta_{n,s}$. \label{tb:comp}}
\end{table*}

\section{GL and BH shadow size in the SFL\label{sec:glbhsz}}

GL in the SFL (and also the weak field limit), including the image apparent angular positions and their magnifications, are always solved from the GL equation that is applicable to the case. Previously, GL equation was usually built using some approximate geometrical relations between the source and lens distances, the source's angular positions, and the deflection angle for source and detector located at infinity. However, in simple spacetimes such as SSS spacetimes, using $\Delta \phi$ that works for source and detector located at finite distances, we show now that an easier and more exact GL equation system can be established. This system contains two equations and is applicable to GL in both strong and weak field limits. It also involves very little geometrical approximation and is usually easy to solve for the image apparent angles.

\begin{figure}
\includegraphics[width=0.4\textwidth]{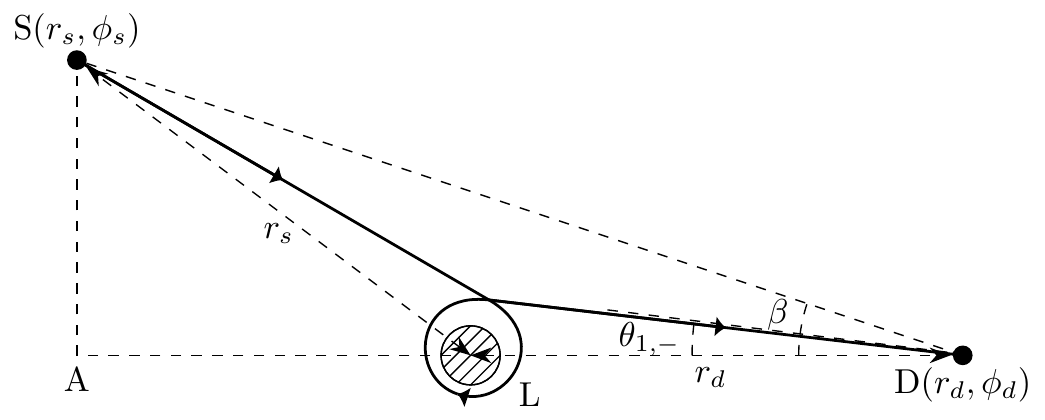}
\caption{The GL in the SFL. The trajectory looped one circle clockwisely. $(r_s,~\phi_s)$ and $(r_d,~\phi_d)$ are the locations of the source S and detector D respectively. L is the lens. $\theta_{1,-}$ is the apparent angle of this particular image. \label{fig:slills}
}
\end{figure}

The first equation links $\Delta\phi$ to the source coordinate $(r_s,~\phi_s)$ and the detector coordinates $(r_d,~\phi_d)$ (see Fig. \ref{fig:slills}). Note that $\phi_s,~\phi_d\in[0,2\pi)$ and without losing any generality, later on we can set $\phi_d=0$. This equation is nothing but the very definition of $\Delta\phi$
\be
\Delta\phi(r_s,~r_d,~b,~p)-2n\pi =\pi(1+s)-s(\phi_s-\phi_d),\label{eq:leq1}
\ee
where integer $n\geq 1$ is the number that the trajectory loops around the center and $p$ is used to denote collectively all other parameters of the spacetime. The $s=\pm 1$ is a sign corresponding to the anti-clockwise and clockwise looping directions respectively. From this equation, one is able to solve for each $n$ the two $b$'s from the two directions that allow the trajectory starting from the source to be received at the detector. Using the $\Delta\phi$ to the $\mathcal{O}(a)^0$ order, which are given in Eqs. \eqref{eq:alphacdef}, Eq. \eqref{eq:leq1} becomes
\be
-C_0\ln \lb 1-\frac{b_c}{b}\rb +D_0-2n\pi =\pi(1+s)-s\Delta\phi_{sd},\ee
where we denoted $\Delta\phi_{sd}=\phi_s-\phi_d$.
Immediately the impact parameter can be solved as
\bea
b_{n,s}&=&\frac{b_c}{\displaystyle  1-\exp\lcb \frac {-\lsb 2 n+(1+s)\rsb\pi+ D_0+s\Delta\phi_{sd}}{C_0}\rcb },\nn\\
&&~~~~(s=\pm 1, ~n= 1,2,3,\cdots) \label{eq:bsol}
\eea
where $b_c$ is given in Eq. \eqref{eq:bcdef}.
Clearly, as $n$ varies, there will be a discrete series of $b$ allowing the signal to reach the detector. Moreover, because the $\Delta\phi(r_s,~r_d,~b,~p)$ we substituted into Eq. \eqref{eq:leq1} become more accurate when $b$ approaches $b_c$, i.e., $a\to 0^+$, the solution \eqref{eq:bsol} will be more precise when $n$ is larger. Note that $b_{n,s}$ are also dependant on the finite distances of source and detector through the coefficient $D_0(\eta_s,\eta_d)$ with $\eta_{s,d}$ given in Eq. \eqref{eq:etedef}, although this dependence is weak especially when $r_s,~r_d$ and $n$ are large.

The purpose of the GL is usually to solve the image apparent angles $\theta$ against the detector-lens axis (see Fig. \ref{fig:slills}) and their magnification. For geodesics in SSS spacetimes, this can be achieved by using an exact apparent angle formula  once $b$ is known
\bea 
\theta_{n,s}& =&\arcsin \frac{b_{n,s}\sqrt{E^2-\kappa}}{\sqrt{\lb \frac{E^2}{A(r_d)}-\kappa\rb C(r_d)}} ,\nn\\
&&~~~~(s=\pm 1, ~n= 1,2,3,\cdots). \label{eq:thetanres}
\eea
We refer the reader to Ref. \cite{ourletter,bk:gr2} for the derivation of this formula. 
Substituting Eq. \eqref{eq:bsol} into the above equation, it is immediately clear that one discrete series of $\theta_{n,s}~(n=1,2,3,\cdots)$ is allowed  from each side ($s=\pm1$) of the lens
\bea
&&\theta_{n,s}=\arcsin \left[\sqrt{\frac{\lb \frac{E^2}{A(r_c)}-\kappa\rb C(r_c)}{\lb \frac{E^2}{A(r_d)}-\kappa\rb C(r_d)}} \nn\right.\\
&&\times\left. \frac{1}{\displaystyle  1-\exp\lcb\frac {-\lsb 2 n+(1+s)\rsb\pi+ D_0+s\Delta\phi_{sd}}{C_0}\rcb }\right].
\label{eq:thetansreexp}
\eea
These are the apparent angular locations of the relativistic images in the SFL.  Here in Eq. \eqref{eq:thetansreexp} for the first time we give an explicit formula for the apparent angle that works for all SSS spacetimes, particle velocity, source and detector distance. An approximate form of this result for null rays was obtained by Bozza in Ref. \cite{Bozza:2002zj} and used/re-derived by many others in particular spacetimes \cite{Perlick:2004tq,Whisker:2004gq, Bozza:2005tg,Eiroa:2005ag,Nandi:2006ds, Chen:2009eu,Stefanov:2010xz,Tsukamoto:2012xs, Sahu:2012er, Wei:2014dka,Pang:2018jpm}. When $n\to\infty$, $\theta_{\infty,s}$ becomes independent of $C_0,~D_0$ as well as $\Delta\phi_{sd}$ and reaches its minimal value, i.e., the size of the black hole shadow
\be
\theta_{\text{Shadow}}=\theta_{\infty,s}=\arcsin \sqrt{\frac{\lb \frac{E^2}{A(r_c)}-\kappa\rb C(r_c)}{\lb \frac{E^2}{A(r_d)}-\kappa\rb C(r_d)}} .
\label{eq:shadowsize}
\ee
It is seen that the shadow size is determined completely by the location of the detector $r_d$, the particle sphere size $r_c$, the metric functions at these locations and the velocity of the signals forming the shadow. Although Eq. \eqref{eq:shadowsize} is not a difficult result, to our best knowledge, such simple relation between the spacetime metric and shadow size was never reported before, certainly not for signals with general asymptotic velocities. 

With the apparent angles of the images known, now we show that the magnifications of the relativistic images form a geometrical series with a convergence factor determined by $C_0$ only. The magnification of the $n$-th relativistic image is defined as \cite{Eiroa:2003jf,Pang:2018jpm}
\be \label{eq:mudef}
\mu_{n,s}=\left|\frac{\theta_{n,s}}{\beta}\frac{\dd\theta_{n,s}}{\dd\beta}\right|,
\ee
where $\beta$ is the angle of the source against the lens-detector axis if there was no GL (see Fig. \ref{fig:slills}). To facilitate the computation of $\mu_{n,s}$, we have to link $\beta$ to other known quantities. This can be done through the geometrical relation in triangle $\triangle SAD$ that
\be
\lsb r_d-r_s\cos\Delta\phi_{sd}\rsb \tan\beta=r_s\sin\Delta\phi_{sd}, \label{eq:betaphisd}
\ee
which links $\beta$ to $\Delta\phi_{sd}$.
Therefore using Eq. \eqref{eq:betaphisd} in Eq. \eqref{eq:mudef} and together with Eq. \eqref{eq:thetansreexp}, the magnification for the $n$-th image under the condition that $r_d\gg b_c$ becomes
\begin{widetext}
\bea
\mu_{n,s}&=&\left|\frac{ \theta_{n,s}}{ \beta}\frac{\dd\theta_{n,s}}{\dd\beta}\right| =\left|\frac{ \theta_{n,s}}{ \beta}\frac{\dd\theta_{n,s}}{\dd\Delta\phi_{sd}} \frac{\dd\Delta\phi_{sd}}{\dd\beta}\right|\nn\\
&\approx &\frac{1}{C_0} \frac{b_c^2}{r_sr_d^3}\frac{\lsb  r_d^2+r_s^2-2r_dr_s\cos\Delta\phi_{sd}\rsb^{3/2}}{\displaystyle \left|\lb\cos \Delta\phi_{sd}-\frac{r_s}{r_d}\rb \arctan\frac{\sin \Delta\phi_{sd}}{\cos \Delta\phi_{sd}-r_d/r_s} \right|} \exp\lcb\frac {-\lsb 2 n+(1+s)\rsb\pi+ D_0+s\Delta\phi_{sd}}{C_0}\rcb. \label{eq:mures}
\eea
\end{widetext}
This magnification depends on $n$ and $\Delta\phi_{sd}$ in some interesting ways. Clearly, as $n$ increases, the magnifications forms a geometric series with the convergence factor $\text{e}^{-\frac{2\pi}{C_0}}$. This means that the relativistic images will become weaker and weaker as one goes to inner images with smaller $\theta_{n,s}$. On the other head, if one holds $n$ and varies $\Delta \phi_{sd}$, then the denominator of Eq. \eqref{eq:mures} indicates that the magnification diverges at
\begin{align}
&(1)~\Delta\phi_{sd}=0,\pi,\arccos\frac{r_s}{r_d},2\pi-\arccos\frac{r_s}{r_d} &&\text{if}~r_s<r_d;\nn\\
&(2)~\Delta\phi_{sd}=\pi &&\text{if}~ r_s=r_d;\label{eq:mudivcases}\\
&(3)~\Delta\phi_{sd}=0,~\pi&&\text{if}~ r_s>r_d.\nn
\end{align}
(see Fig. \ref{fig:maghwd}).
Moreover, because $0\leq\Delta\phi_{sd}<2\pi$, one can also verify from Eq. \eqref{eq:mures} that the magnification of the $n$-th relativistic image with  $\Delta\phi_{sd}$ that is not close to the detector-lens axis is always larger than
that of the $(n+1)$-th image with any $\Delta \phi_{sd}$ which might be close to $0$ or $\pi$. This implies that when discussing series of relativstic images in GL in the SFL, the angular location of the source is not limited to $\Delta\phi \approx \pi$ as in the case of GL in the weak field limit or $\Delta\phi\approx 0$ as in the case of retrolensing.

Since usually the relativistic images with $n\geq 2$ are not resolvable from each other in reasonable future (see Sec. \ref{sec:rn}), the images with $n\geq 2$ will only form one image. The total magnification of this image is
\bea
\mu_{\sum,s}&=&\sum_{n=2}^\infty\mu_{n,s}\nn\\
&=&\mu_{1,s} \lsb\exp\lb\frac{2\pi}{C_0}\rb-1\rsb^{-1}.
\eea

To deduce properties of the BH spacetime, which are characterized by $C_0$ and $D_0$, we can use the flux ratio of this images against the first relativistic image \cite{Bozza:2002zj}, which is
\be \label{eq:rmudef}
r_\mu\equiv 	\frac{\mu_{1,s}}{\mu_{\sum,s}}= \exp\lb\frac{2\pi}{C_0}\rb-1.
\ee
Moreover, one can also use the positions of $\theta_{1,s}$ and $\theta_{\infty,s}$ to find two ratios
\bea
r_{\theta,s}&=&\left|\frac{\theta_{1,s}-\theta_{\infty,s}}{\theta_{\infty,s}}\right|\nn\\
&\approx&\exp\lsb\frac {-(3+s)\pi+ D_0+s\Delta\phi_{sd}}{C_0}\rsb ~(s=\pm1).\label{eq:rthetadef}
\eea
From \eqref{eq:rmudef} and \eqref{eq:rthetadef}, one is able to solve the coefficients $C_0$ and $D_0$, and the angular coordinate difference $\Delta\phi_{sd}$
\bea
&&C_0=\frac{2\pi}{\ln \lb r_{\mu}+1\rb},\\
&&D_0=\frac{\pi\ln\lsb \lb r_\mu+1\rb^3r_{\theta,+}r_{\theta,-}\rsb}{\ln \lb r_{\mu}+1\rb},\\
&&\Delta\phi_{sd}=\frac{\pi\ln\lsb \lb r_\mu+1\rb r_{\theta,+}/r_{\theta,-}\rsb}{\ln \lb r_{\mu}+1\rb}.
\eea
Again, as pointed out by Bozza \cite{Bozza:2002zj}, by measuring the flux and angular position ratios, one is able to reconstruct the full SFL expansion of the deflection angle. Further more, since $C_0$ and $D_0$ depend on the nature of the BH, this also allow us to deduce information about the central BH. 

\section{The Hayward BH spacetime case \label{sec:rn}}

In this section, we apply our perturbative method for solving $\Delta\phi$ and the GL equations in the SFL to the Hayward BH spacetime, whose metric is given by \cite{Hayward:2005gi}
\be
\label{eq:hwdmetric}
A(r)=1-\frac{2m r^2}{r^3+2l^2m},~B(r)=\frac{1}{A(r)},~C(r)=r^2,
\ee
where $m$ is the spacetime mass and the charge parameter $l$ is smaller than a critical value $l<4m/(3\sqrt{3})\equiv l_c$ in order for the spacetime to contain a BH.
The equation determining the location of the particle sphere is given by Eq. \eqref{eq:rcdef}, which for the Hayward BH spacetime becomes
\bea
&&0=\left(E^2-\kappa \right)r_c^6
+m \left(4 \kappa -3 E^2\right)r_c^5-4 \kappa  m^2r_c^4 \nn\\
&&+4 l^2 m \left(E^2-\kappa \right) r_c^3+8 \kappa  l^2 m^2 r_c^2+4 l^4 m^2 \left(E^2-\kappa \right).
\eea
After solving this equation numerically for a given $E$ (or $v$), $m$ and $l$ then using Eq. \eqref{eq:bcdef}, one can obtain the corresponding critical impact parameter.

\begin{figure}
\includegraphics[width=0.4\textwidth]{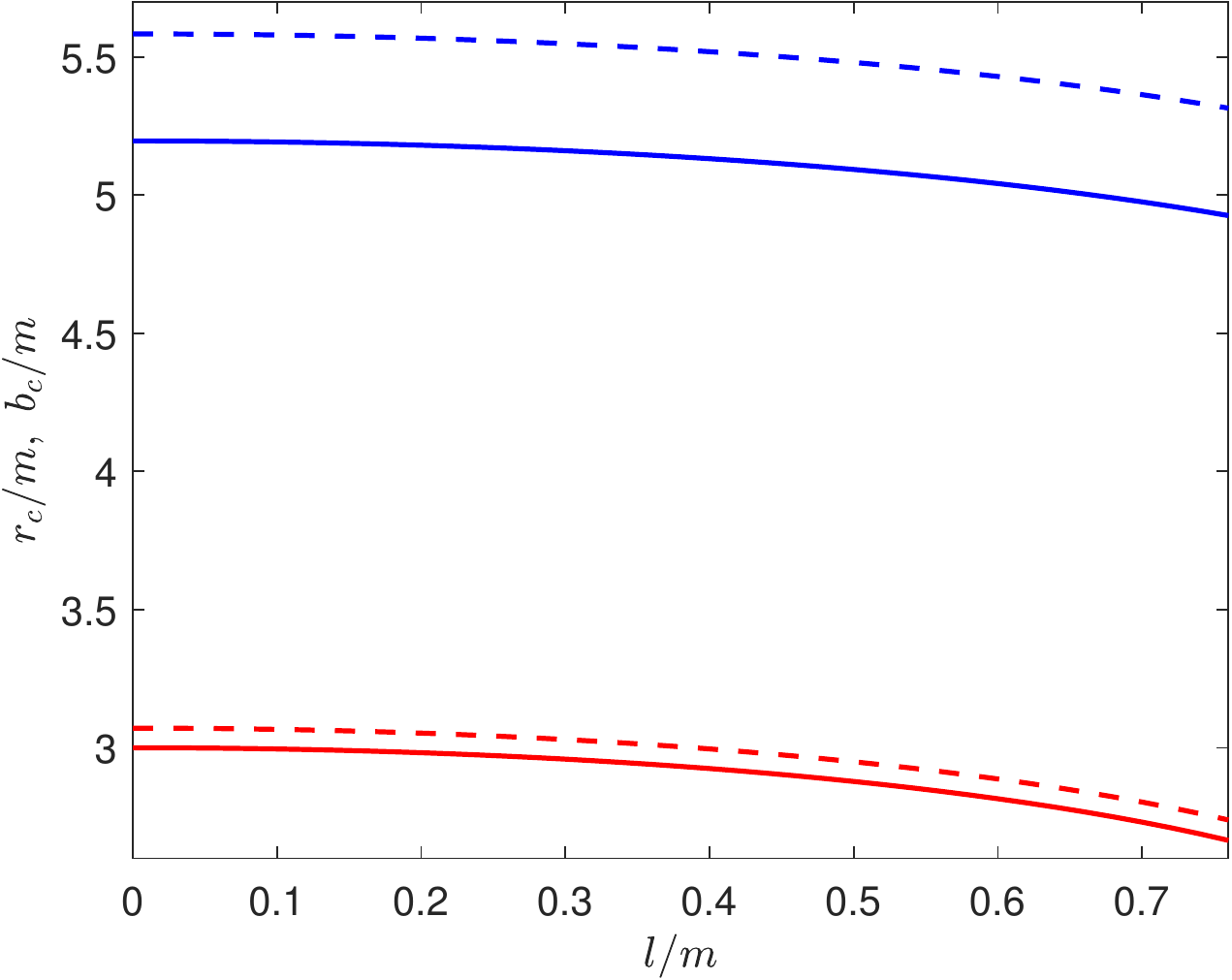}
\caption{\label{fig:bcr0res} The critical radius $r_c$ (red) and impact parameter $b_c$ (blue) as functions of $l$ for the Hayward BH spacetime for $v=1$ (solid) and $v=0.9$ (dashed) respectively. }
\end{figure}

In Fig. \ref{fig:bcr0res}, we plot the solved $r_c$ and $b_c$ as a function of $l$ from 0 to $l_c$ for two different velocities. It is seen that the with the increase of $l$, both $r_c$ and $b_c$ decrease monotonically from their Schwarzschild values, i.e. $3m$ and $3\sqrt{3}m$. Then one would expect that the gravitational effect become stronger if the test particles come close to the same distance to the critical sphere. The decrease of the signal velocity increases $b_c$ more dramatically than $r_c$, which is also expected because it is known that in Schwarszchild spacetime as $v$ decreases to zero, $b_c$ approaches infinity while $r_c$ only increases to $4m$. Similar increase of $b_c$ and $r_c$ also happens in the RN BH spacetime \cite{Pang:2018jpm}.

Once $r_c$ is known, then the expansions of the metric functions in \eqref{eq:hwdmetric} around $r_c$ can be worked out. To the $\mathcal{O}(r-r_c)^4$, they are
\begin{widetext}
\begin{subequations}\label{eq:rnbhexp}
\begin{align}
    A(r)=&\left(1-\frac{2 m r_c^2}{r_c^3+2 l^2 m}\right)
    +\frac{2 m r_c \left(r_c^3-4 l^2 m\right)}{\left(r_c^3+2 l^2 m\right)^2}\left(r-r_c\right) 
    -\frac{2 m \left(-14 l^2 m r_c^3+r_c^6+4 l^4 m^2\right)}{\left(r_c^3+2 l^2 m\right)^3} \left(r-r_c\right)^2
    \nonumber\\&
    +\frac{2 m  \left(40 l^4 m^2 r_c^2-32 l^2 m r_c^5+r_c^8\right)}{\left(r_c^3+2 l^2 m\right)^4}\left(r-r_c\right)^3
    +\mathcal{O}\left(r-r_c\right)^4\\
    B(r)=
    &\frac{1}{1-\frac{2 m r_c^2}{r_c^3+2 l^2 m}}+\frac{2 m r_c \left(4 l^2 m-r_c^3\right)}{\left(-2 m r_c^2+r_c^3+2 l^2 m\right)^2}\left(r-r_c\right) 
    +\frac{2 m  \left(12 l^2 m^2 r_c^2-14 l^2 m r_c^3+r_c^6+4 l^4 m^2\right)}{\left(-2 m r_c^2+r_c^3+2 l^2 m\right)^3}\left(r-r_c\right)^2
    \nonumber\\&
    +\frac{2 m r_c  \left[32 l^4 m^3-r_c \left(-32 l^2 m^3 r_c+56 l^2 m^2 r_c^2-32 l^2 m r_c^3+r_c^6+40 l^4 m^2\right)\right]}{\left(-2 m r_c^2+r_c^3+2 l^2 m\right)^4}\left(r-r_c\right)^3
    +\mathcal{O}\left(r-r_c\right)^4\\
C(r)=&r_c^2+2r_c(r-r_c)+(r-r_c)^2.
\end{align}
\end{subequations}
Reading off the corresponding coefficients $a_n,~b_n$ and $c_n$ and substituting into Eqs. \eqref{eq:y0nres}, one can obtain the coefficients in the deflection angle of the Hayward spacetime
\begin{subequations}\label{eq:ynexprnbh}
\begin{align}
    &y_{-1,0,\mathrm{H}}=\frac{\left(r_c^3+2 l^2 m\right){}^{3/2} }{r_c} \frac{\lb -2 m r_c^2+r_c^3+2 l^2 m\rb^{1/2}}{\lb -96 l^4 m^3 r_c^3+96 l^2 m^3 r_c^5-48 l^2 m^2 r_c^6-6 m^2 r_c^8+4 m r_c^9+32 l^6 m^4\rb^{1/2}},\\
    &y_{0,0,\mathrm{H}}=\frac{2 r_c \left(r_c^3+2 l^2 m\right){}^{3/2} \left(-2 m r_c^2+r_c^3+2 l^2 m\right){}^{5/2} \left(-32 l^2 m r_c^3+r_c^6+40 l^4 m^2\right)}{m \left(-48 l^4 m^2 r_c^3+48 l^2 m^2 r_c^5-24 l^2 m r_c^6-3 m r_c^8+2 r_c^9+16 l^6 m^3\right){}^2}.
\end{align}
\end{subequations}
Higher order terms can be similarly obtained but they are too long to presented here.
Substituting them further into Eqs. \eqref{eq:alpharesordera0}, the deflection angle in the Hayward BH spacetime in the SFL becomes
\bea
\Delta\phi_\text{H}&\approx& -\sqrt{2}y_{-1,0,\mathrm{H}}\ln \lb 1-\frac{b_c}{b}\rb \nn\\
&&+\sqrt{2}y_{-1,0,\mathrm{H}}\ln \lb 2\sqrt{\eta_s\eta_d}\rb+\sqrt{2}y_{0,0,\mathrm{H}}\lb \sqrt{\eta_s}+\sqrt{\eta_d}\rb+\sum_{n=1}^\infty \frac{2y_{n,0}  \eta_i^{\frac{n+1}{2}}} {2^{\lsb\frac{n+1}{2}\rsb+\frac12}(n+1)}\nn\\
&&+~\text{higher order terms}. \label{eq:dphdef}
\eea
\end{widetext}

\begin{figure}[htp!]
\includegraphics[width=0.4\textwidth]{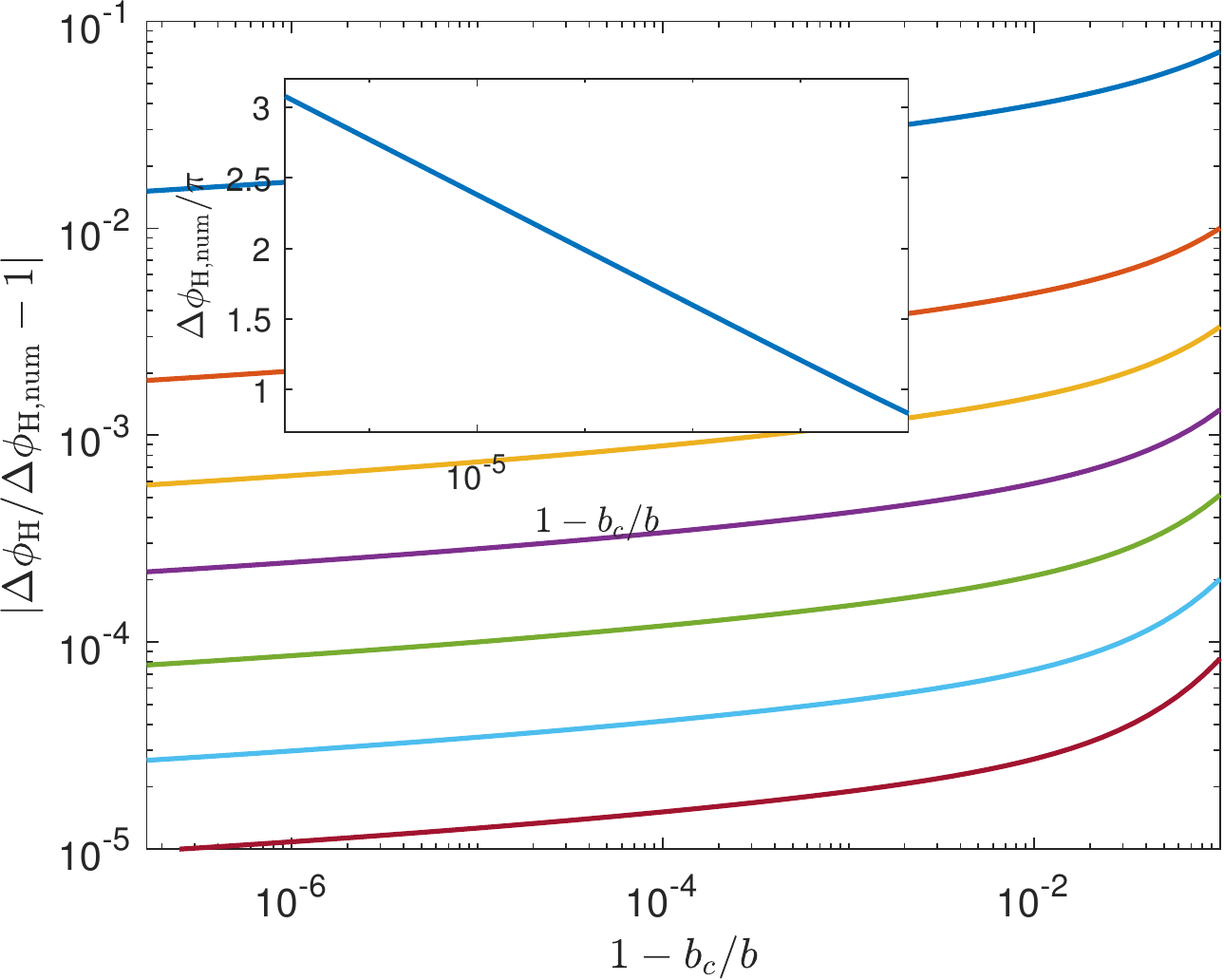}\\
(a)\\
\includegraphics[width=0.4\textwidth]{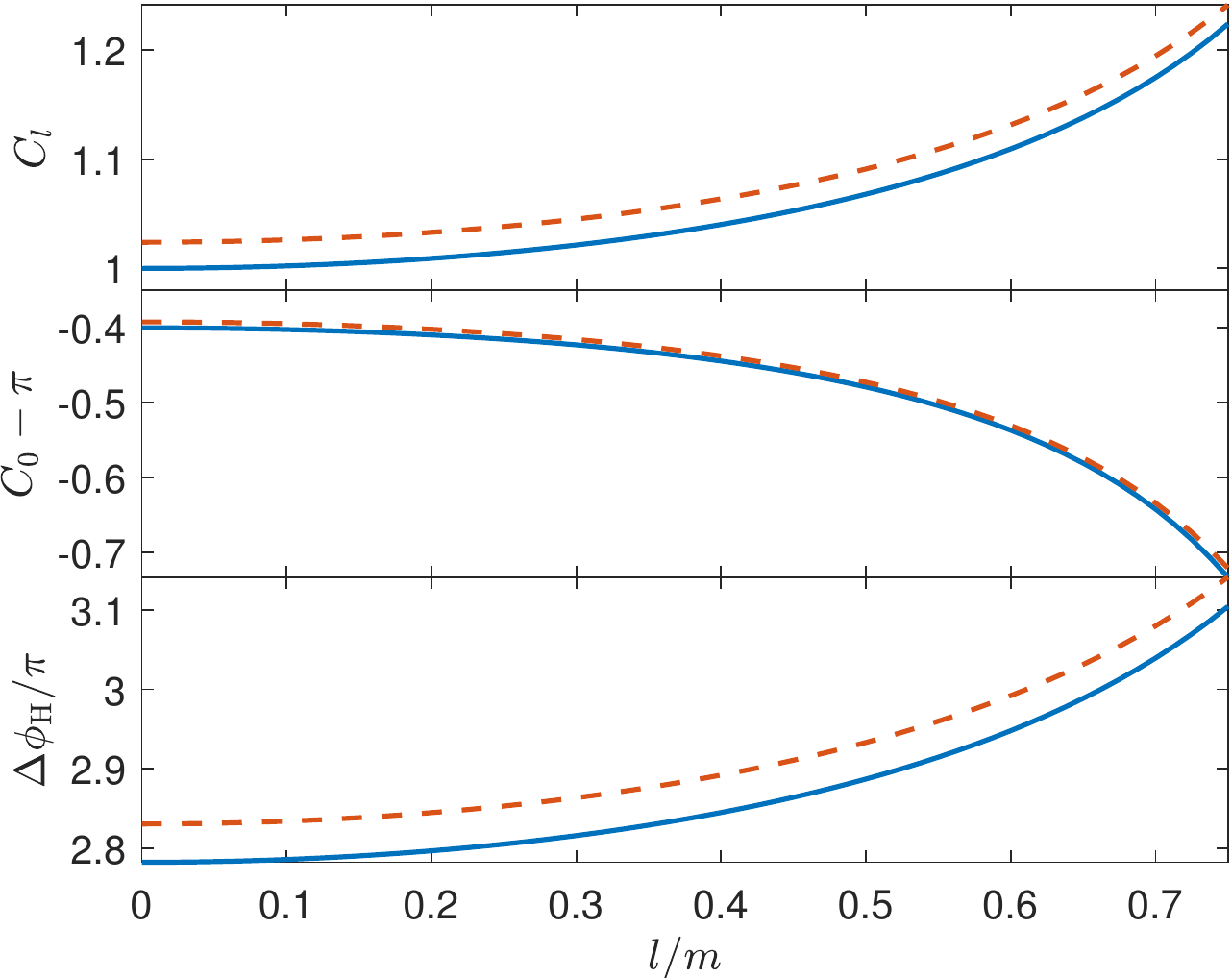}\\
(b)
\caption{\label{fig:alphahwd} (a) $|\Delta \phi_{\text{H}}/\Delta\phi_{\text{H,num}}-1|$ for $l/m=1/2$ and $v=1$ and  Eq. \eqref{eq:dphdef}  truncated at order 0, 2, 4, 6, 8, 10 and 12 (from bottom to top). Inset: numerical value obtained from Eq. \eqref{eq:deflection}.  (b) Variation of coefficients $C_0,~D_0$ and deflection $\Delta\phi_{\text{H}}$ for the Hayward spacetime with respect to $l/m$ for velocities $v=1$ (solid) and $v=0.9$ (dashed) respectively.  }
\end{figure}

In Fig. \ref{fig:alphahwd} (a), we plot the relative difference $|\Delta \phi_{\text{H}}/\Delta\phi_{\text{{H,num}}}-1|$ between $\Delta\phi_{\text{H}}$ obtained using Eq. \eqref{eq:dphdef} with the truncation of the summation to order 0, 2, 4, 6, 8, 10 and 12 (from bottom to top) and the corresponding numerical value $\Delta\phi_\text{H,num}$ obtained using numerical integration of the definition Eq. \eqref{eq:deflection} as functions of $b$ as it approaches $b_c$ from above. It is seen that as the order of truncation in the summation of Eq. \eqref{eq:dphdef} increases, the deviation of our analytical result from the true value obtained by the numerical integration of Eq. \eqref{eq:deflection} decreases, by a factor of $\sim 3$ for every two orders. At order 12, the relative deviation is only about $10^{-5}$ when $1-b_c/b$ is about $10^{-7}$, which corresponds to the $\Delta\phi_{\text{H}}\approx 3\pi$, i.e., the trajectory loops 1 circle around the BH.  Moreover, for each fixed order, the deviation decreases monotonically as $b$ approaches $b_c$. Both the above features match the fact that our result \eqref{eq:dphdef} is essentially a perturbative series approximating the true $\Delta\phi$. 

In Fig. \ref{fig:alphahwd} (b), the coefficients $C_0,~D_0$ and $\Delta\phi_{\text{H}}$ as functions of the parameter $l$ for different values of $v$ are plotted. The increase of $C_0$ and decrease of $D_0$ as the charge parameter $l$ increases are very similar to the effect of electric charge $Q$ in the case of deflection in the SFL in RN BH spacetime \cite{Pang:2018jpm}. This plot also shows that the $\Delta\phi_{\text{H}}$ increases as $l$ increases, which is in accordance to the decrease of $b_c$ in Fig. \ref{fig:bcr0res}. On the other hand,  although the decrease of the velocity from 1 to 0.9 in this case  still increases $\Delta\phi_{\text{H}}$, its effect is much weaker than the increase of $l$, unlike the case in Fig. \ref{fig:bcr0res}. This is a reflection that the decrease of  $v$ indeed increases the particle sphere size and therefore the looping of the trajectory happens at a larger radius, reducing  $\Delta\phi_\text{H}$'s amount of increase.  

\begin{figure}[htp!]
\includegraphics[width=0.4\textwidth]{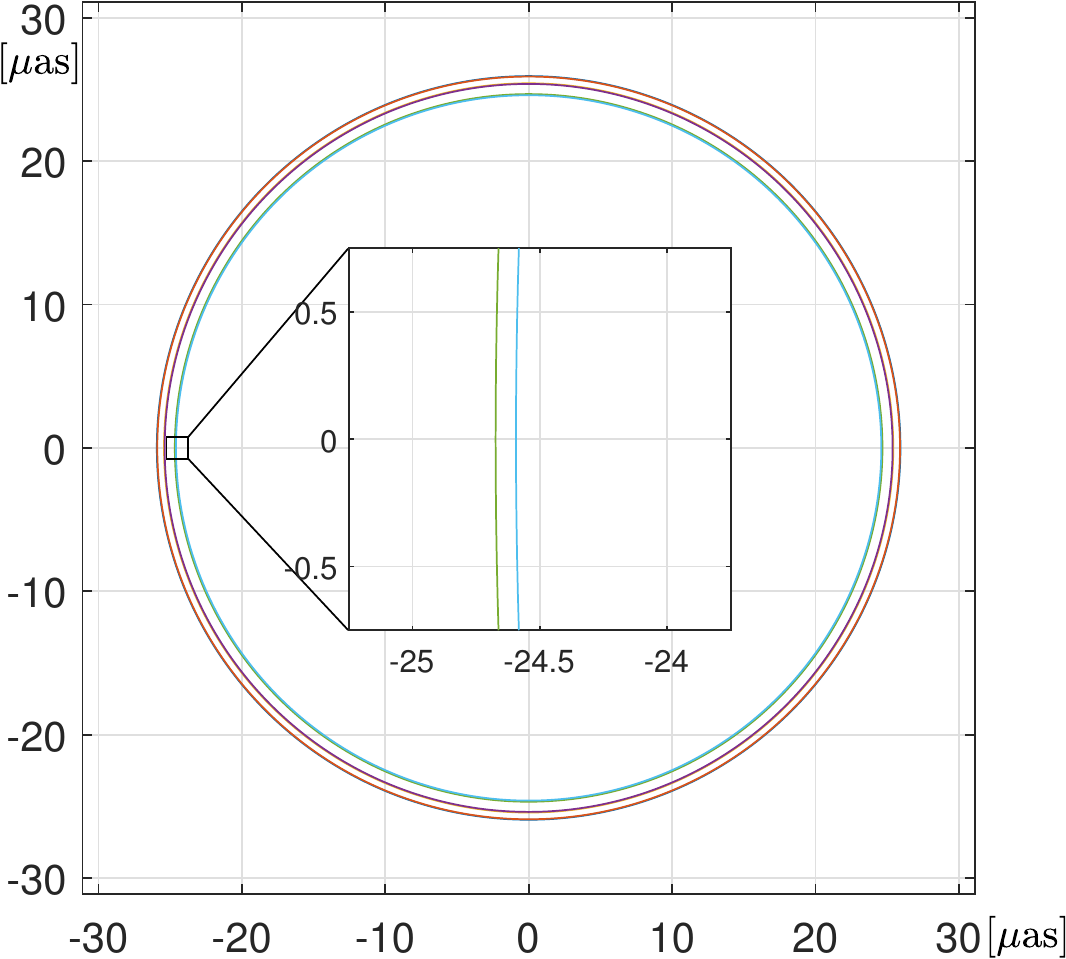}\\
(a)\\
\includegraphics[width=0.4\textwidth]{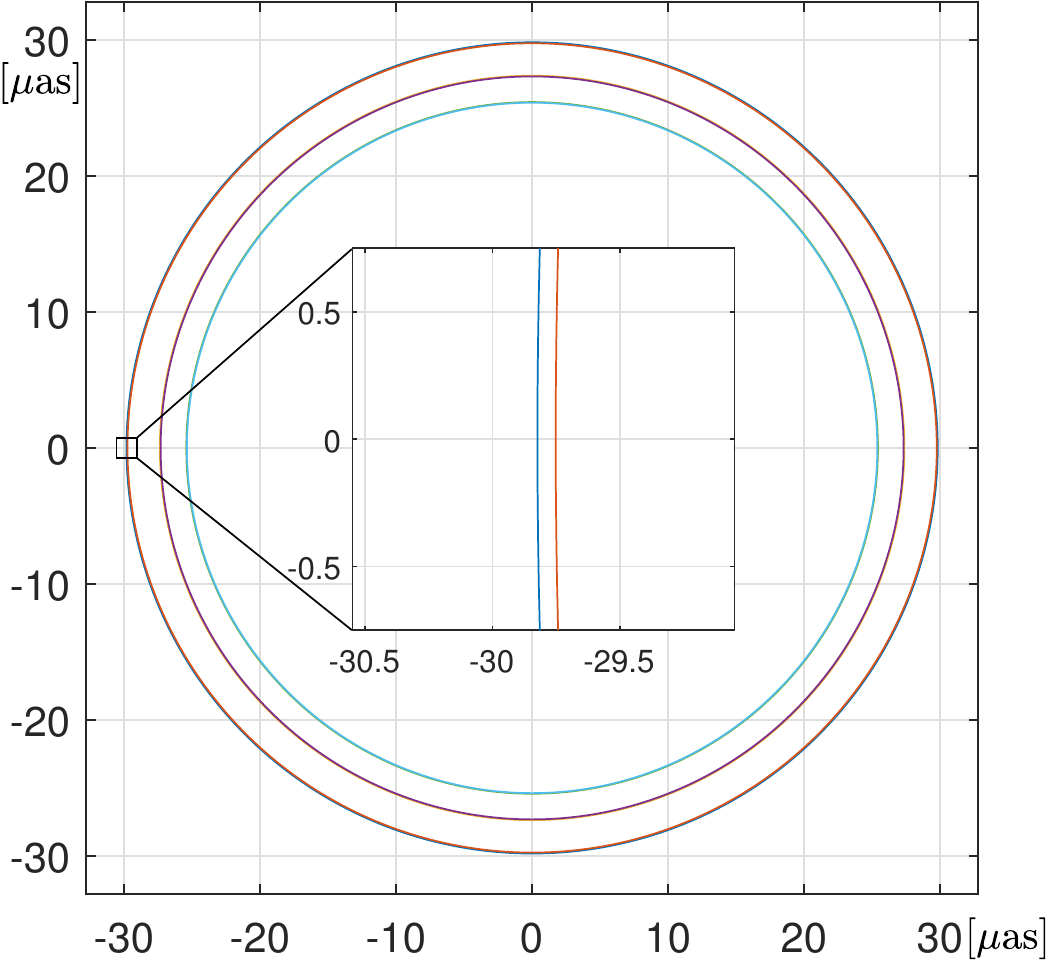}\\
(b)\\
\includegraphics[width=0.4\textwidth]{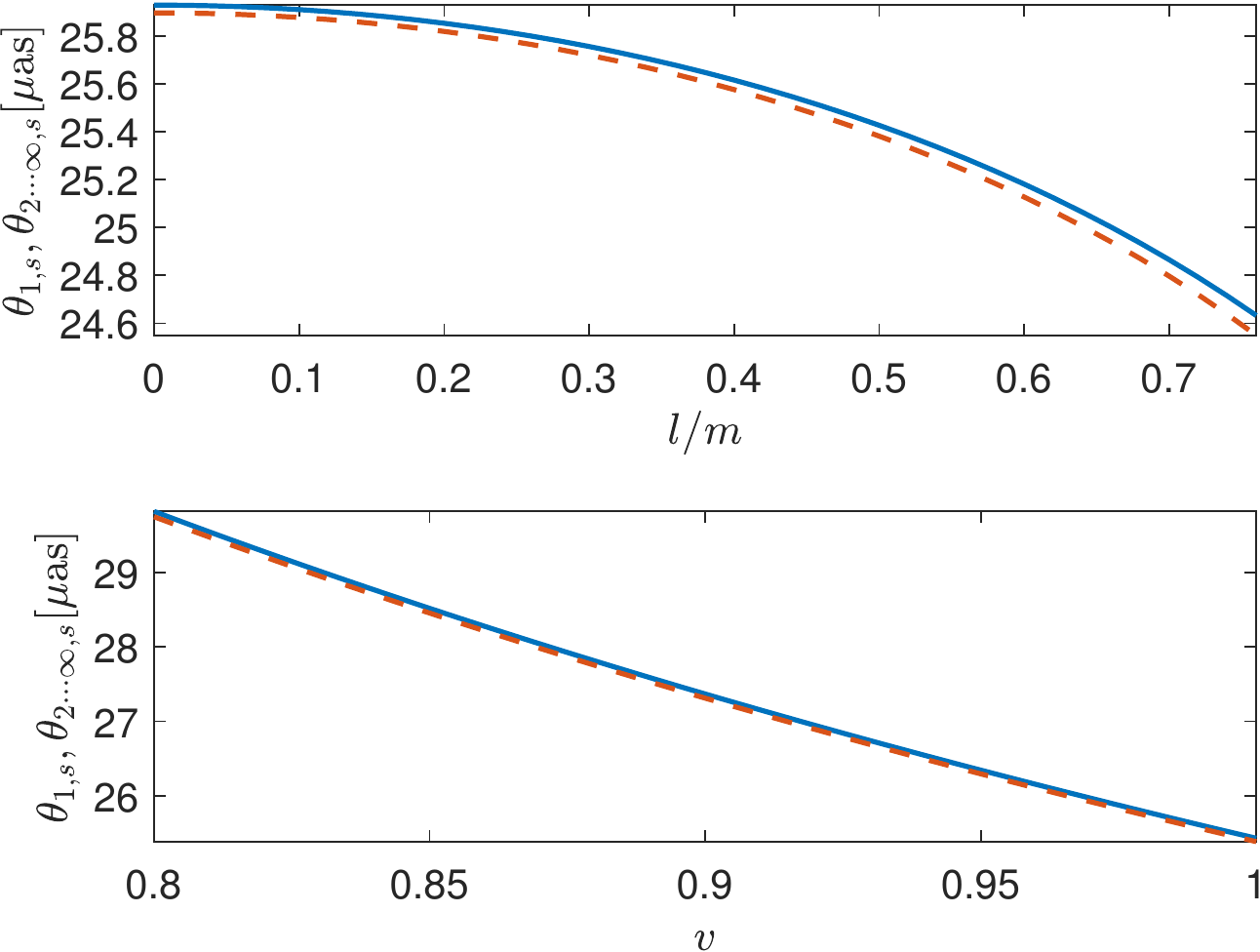}\\
(c)
\caption{\label{fig:thetansh} (a) Relativistic Einstein rings  assuming that the SgrA* BH is an Hayward BH for $v=1$ and $l/m=0, ~1/2,~3/4$ from outer to inner circle. Zoom-in: Partially resolved $\theta_{1,s}$ and $\theta_{2\cdots\infty,s}$.  (b) Relativistic Einstein rings for $l/m=1/2$ and $v=1,~0.9, ~0.8$ from inner to outer respectively. Zoom-in: Partically resolved Einstein rings for $v=0.8$. (c) The dependence of $\theta_{1,s}$ and $\theta_{\infty,s}$ on (top): $l$ while holding $v=1$ and (bottom): $v$ while holding $l/m=0.5$. }
\end{figure}

Substituting the $\Delta\phi_{\text{H}}$ in Eq. \eqref{eq:dphdef} into Eq. \eqref{eq:thetansreexp}, we are able to solve the apparent angles $\theta_{n,s}$. In Fig. \ref{fig:thetansh} (a) and (b), we plot the Einstein rings in the SFL, which are located at $\theta_{n,s}$ when $\Delta\phi_{sd}=\pi$. Here we assume that $m=4.12\times 10^6M_\odot$ is the SgrA* BH mass, $r_s=r_d=8.12$ [kpc] is the distance to SgrA* BH \cite{Andreas:2018}. However, note that in the zoom-out large plot, different rings correspond to the relativistic Einstein rings of different $l$ (from outer to inner circle in Fig. \ref{fig:thetansh} (a), $l/m=0, ~1/2,~3/4$) or different $v$ (from inner to outer circle in Fig. \ref{fig:thetansh} (b), $v=1,~0.9,~0.8$ respectively). Only in the zoom-in insets the Einstein rings for a fixed $l$ or $v$ are resolved between $\theta_{1,s}$ and $\theta_{2\cdots \infty,s}$. In general, the first relativistic ring  $\theta_{1,s}$ is always larger than higher order ones $\theta_{2\cdots\infty,s}$, separated from them by roughly $\mathcal{O}(0.1)$ [$\mu$as]. Therefore, these higher order rings are not even observationally resolvable in the near future. Their locations are practically the same as the location of the BH shadow of the Hayward BH. In Fig. \ref{fig:thetansh} (c) top and bottom plots respectively, the dependence of the outmost ring location $\theta_{1,s}$ and the inner overlapped ring location $\theta_{2\cdots \infty,s}$ on the spacetime parameter $l$ and signal variable $v$  are plotted. It is clear that as $l$ increases or $v$ decreases, the Einstein ring and shadow sizes decrease or increase respectively. Moreover, change of $v$ has a larger effect than that of $l$ on $\theta_{n,s}$, in agreement with their effects on $b_c$ as indicated  in Fig. \ref{fig:bcr0res}. 
These features in the Hayward spacetime are similar to the effect of charge and velocity in the RN spacetime \cite{Pang:2018jpm}.

\begin{figure}[htp!]
\includegraphics[width=0.4\textwidth]{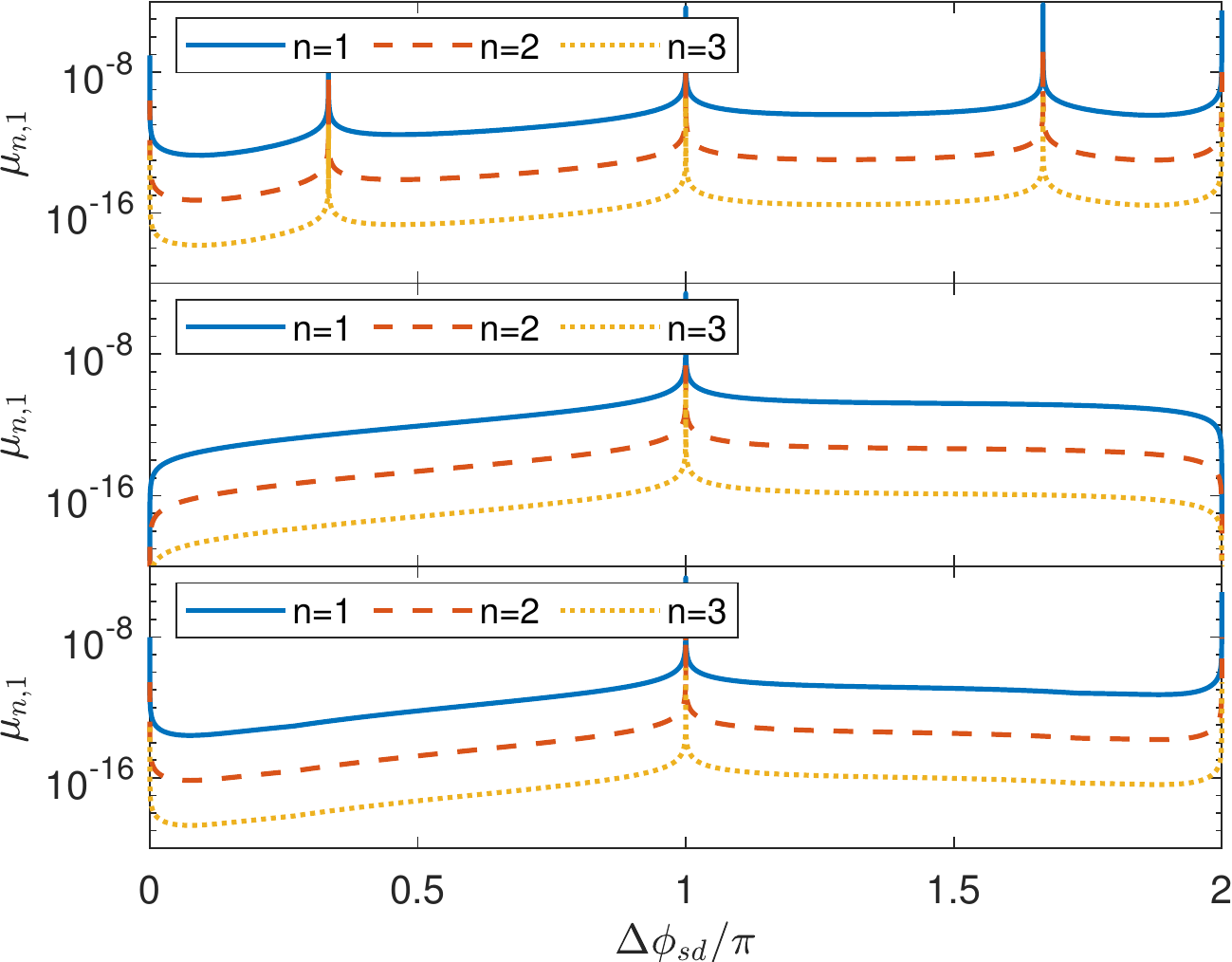}\\
(a)\\
\includegraphics[width=0.4\textwidth]{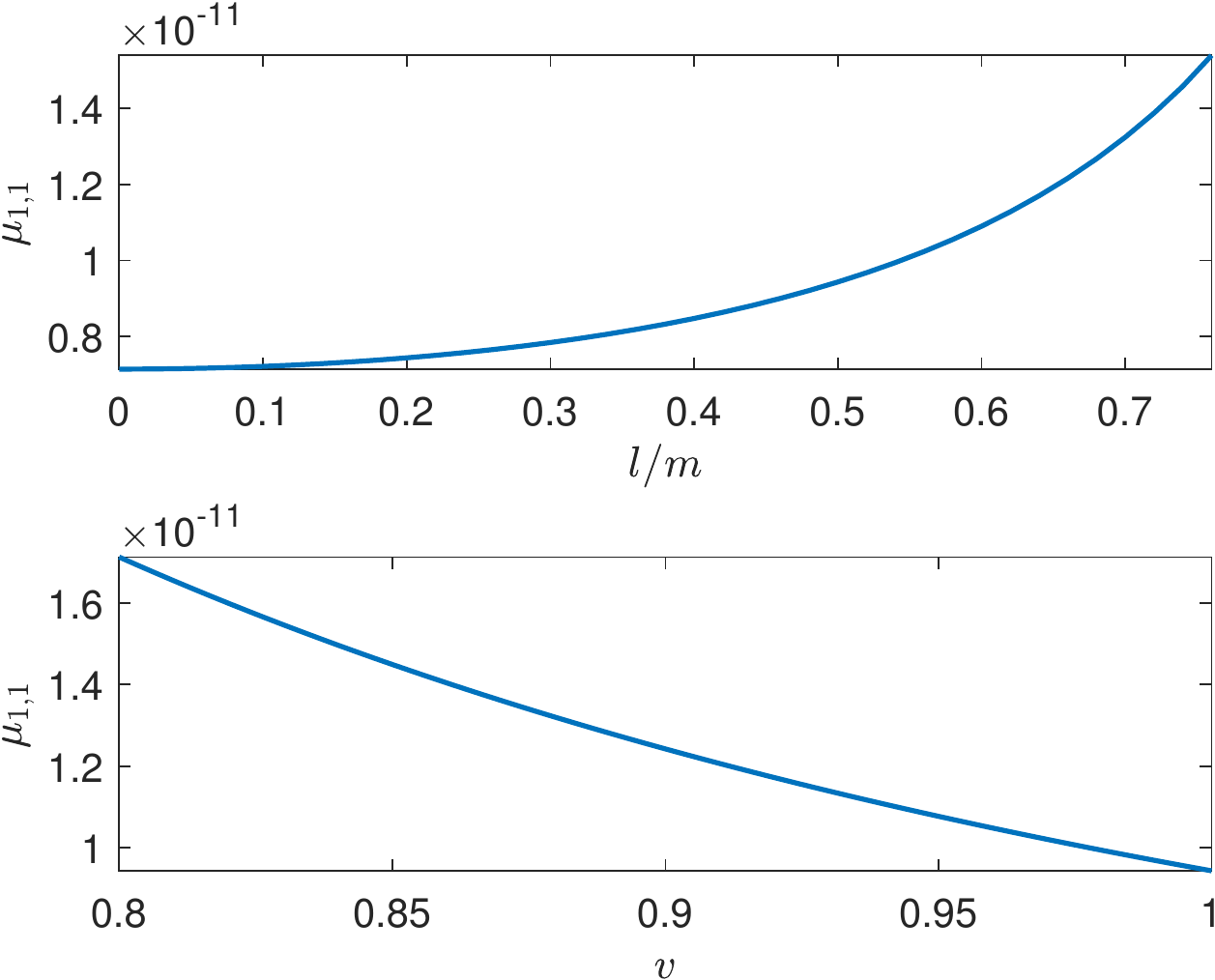}\\
(b)
\caption{\label{fig:maghwd} (a) The magnification of the relativistic images in the Hayward BH spacetime. $m,~r_d,~v,~l$ are the same as in Fig. \ref{fig:thetansh} and $r_s=r_d/2$ (top), $r_s=r_d$ (center) and $r_s=1.5r_d$ (bottom) for $n=1,~2$ and 3. (b) The dependence of the magnification on $l$  while holding $v=1$ (top) and on $v$ while holding $l/m=0.5$ (bottom). $\Delta \phi_{sd}=5\pi/6$ and $r_s=r_d$ is assumed. }
\end{figure}

The magnification of the relativistic images in the Hayward BH spacetime can be obtained using Eq. \eqref{eq:mures}. In Fig. \ref{fig:maghwd} (a), we plot the magnification of the images for different $n$, $\Delta\phi_{sd}$ and $r_s/r_d$, and in Fig. \ref{fig:maghwd} (b) the magnification as a function of $l/m$ and $v$. It is seen from Fig. \ref{fig:maghwd} (a) that as expected, for all $\Delta\phi_{sd}$ and $r_s/r_d$, the magnification $\mu_{1,s}$ of the outer-most image is always much larger than that of the inner images. Because the ratio of magnifications of the two successive images is given by $\exp (-2\pi/C_0)$ and $C_0<2$, $\mu_{1,s}$ is indeed larger than $\mu_{\sum,s}$, as predicted by Eq. \eqref{eq:rmudef}.
The three subplots also verify the three possible cases found in Eq. \eqref{eq:mudivcases}. The existence of these three cases is robust and purely due to the geometric variables $\Delta\phi_{sd},~r_s$ and $r_d$. 
Fig. \ref{fig:maghwd} (b) shows the effect of $l$ and $v$ to the magnification. Given that both the decrease of $v$ and increase of $l$ will increase the value of $C_0$, as can be seen from the Fig. \ref{fig:alphahwd} (b), using Eq. \eqref{eq:mures} then it is immediately clear that these changes will increase the magnification too. This implies that the the Hayward BH with larger charge or signal with slower velocity tends to have larger magnification of the relativistic images, which is confirmed in Fig. \ref{fig:maghwd} (b).

\section{Discussions}

In this work, we developed a perturbative method to compute the deflection angle of both null and timelike signals in the SFL in the SSS spacetimes. In doing so, the effects of finite source/detector distances are also taken into account. The result takes a quasi-power series form of Eq. \eqref{eq:dphgform}, i.e., 
\be
\Delta\phi=\sum_{n=0}\lsb C_n\ln a+D_n\rsb  a^n,
\ee
where $a=\lb 1-\frac{b_c}{b}\rb$ is the infinitesimal expansion parameter as $b\to b_c^+$ and the series coefficients of $a^n$ contain a weak logarithmic divergence. The $a^0$ order for light ray has been well known and widely applied in many spacetimes. The form of the first few orders in Schwarzschild spacetime was also known for light signal by expanding the elliptical functions representing the deflection angle \cite{Iyer:2006cn}.
However, the general form of $\Delta\phi$ for higher order terms for general SSS spacetime, especially for timelike signal is first derived here. 
Using this result, we connected the behavior of the spacetime near the particle (or photon) sphere, the particle velocity and the finite distances to the  deflection angle in the SFL. 

Using a new set of GL equations, we were able to solve the apparent angles and their magnifications of the relativistic images in the SFL for signals with arbitrary velocity. A simple formula for the shadow size was found in terms of the metric functions, particle sphere size and detector radius. 

Applying the perturbative method and results to the Hayward spacetime, we verified that the series approaches the true deflection angle as the truncation order increases and $b\to b_c^+$. As the charge parameter $l$ increases, the particle sphere size decreases, the deflection angle increases, the apparent angle decrease and magnification increases. On the other hand, as the decrease of the signal velocity from that of light, the particle sphere size, the deflection angle, the apparent angle and magnification all increase.

\acknowledgments
We thank Dr. Nan Yang and Mr. Haotian Liu for valuable discussions. This work is supported by the NNSF China 11504276.

\appendix

\section{High order $y_{n,m}$ and integration formulas\label{app:ellp}}

\begin{widetext}
The higher order coefficients of $y_{0,n}$ in Eq. \eqref{eq:y0nres} are
\begin{align}
y_{1,0}=&\frac{T_0^{3/2}}{2\sqrt{2} b_0^{3/2} c_0^{5/2} T_2^{7/2}}(-b_1^2 c_0^2 T_2^2+4 b_0 b_2 c_0^2 T_2^2-2 b_0 b_1 c_0 c_1 T_2^2+3 b_0^2
   c_1^2 T_2^2-4 b_0^2 c_0 c_2 T_2^2+6 b_0^2 c_0^2 T_2^3/T_0-6
   b_0 b_1 c_0^2 T_2 T_3\nn\\
   &
   +6 b_0^2 c_0 c_1 T_2 T_3+15 b_0^2 c_0^2 T_3^2-12
   b_0^2 c_0^2 T_2 T_4),\tag{\ref{eq:y0nres}c}\\
y_{2,0}=&\frac{T_0^2}{4 b_0^{5/2} c_0^{7/2} T_2^5}(
   b_1^3 c_0^3 T_2^3-4 b_0 b_1 b_2 c_0^3 T_2^3+8 b_0^2 b_3 c_0^3 T_2^3+b_0
   b_1^2 c_0^2 c_1 T_2^3-4 b_0^2 b_2 c_0^2 c_1 T_2^3+3 b_0^2 b_1 c_0 c_1^2
   T_2^3-5 b_0^3 c_1^3 T_2^3\nn\\
   &
   -4 b_0^2 b_1 c_0^2 c_2 T_2^3+12 b_0^3 c_0 c_1
   c_2 T_2^3-8 b_0^3 c_0^2 c_3 T_2^3+9 b_0^2 b_1 c_0^3
   T_2^4/T_0-9 b_0^3 c_0^2 c_1 T_2^4/T_0+4 b_0 b_1^2 c_0^3
   T_2^2 T_3-16 b_0^2 b_2 c_0^3 T_2^2 T_3\nn\\
   &
   +8 b_0^2 b_1 c_0^2 c_1 T_2^2
   T_3-12 b_0^3 c_0 c_1^2 T_2^2 T_3+16 b_0^3 c_0^2 c_2 T_2^2 T_3-18
   b_0^3 c_0^3 T_2^3 T_3/T_0+24 b_0^2 b_1 c_0^3 T_2 T_3^2-24 b_0^3
   c_0^2 c_1 T_2 T_3^2-64 b_0^3 c_0^3 T_3^3\nn\\
   &
   -16 b_0^2 b_1 c_0^3 T_2^2 T_4+16
   b_0^3 c_0^2 c_1 T_2^2 T_4+96 b_0^3 c_0^3 T_2 T_3 T_4-32 b_0^3 c_0^3
   T_2^2 T_5)
 \tag{\ref{eq:y0nres}d}
\end{align}
where $T_n$ are given in Eqs. \ref{eq:tnres}
\begin{align}
T_4=&\frac{E^2}{E^2-\kappa}\left[c_4\left(\frac{1}{a_0}-\frac{\kappa}{E^2} \right)
+\frac{\left(a_1^4-3 a_0 a_1^2a_2 +2 a_0^2 a_1a_3 +a_0^2 a_2^2-a_0^3 a_4\right) c_0 }{a_0^5}
+\frac{\left(a_1^2-a_0 a_2\right) c_2 }{a_0^3}-\frac{a_1 c_3}{a_0^2}\right.\nn\\
&\left.-\frac{\left(a_1^3-2 a_0 a_1 a_2+a_0^2 a_3\right) c_1}{a_0^4}\right],\tag{\ref{eq:tnres}c}\\
T_5=&\frac{E^2}{E^2-\kappa}\left[c_5\left(\frac{1}{a_0}-\frac{\kappa}{E^2} \right)+\frac{\left(a_1^4-3 a_0 a_1^2a_2 +a_0^2 a_2^2+2 a_0^2 a_1a_3 -a_0^3 a_4\right) c_1}{a_0^5}+\frac{\left(a_1^2-a_0 a_2\right) c_3}{a_0^3}-\frac{a_1 c_4 }{a_0^2}\right.\nn\\
&
\left.-\frac{\left(a_1^3-2 a_0 a_1 a_2+a_0^2 a_3\right) c_2 }{a_0^4}-\frac{\left(a_1^5-4 a_0 a_1^3a_2 +3 a_0^2  a_1a_2^2+3 a_0^2a_1^2 a_3 -2 a_0^3 a_2 a_3-2 a_0^3 a_1 a_4+a_0^4 a_5\right) c_0 }{a_0^6}\right] \tag{\ref{eq:tnres}d}.
\end{align}

For the integrals appearing in Eq. \eqref{eq:defang4}, we have for non-negative integer $m$ the following result
\begin{subequations}
\label{eq:ttgenintmk}
\begin{align}
n=2k,&\int_a^{\eta_i}\frac{\xi^k}{\sqrt{\xi-a}} \dd \xi=\sum_{j=0}^{k} \frac{2 C_{j}^{k}a^{k-j}\lb \eta_i-a\rb^{j+1/2}}{2j+1}, \label{eq:ttgenintmek} \\
n=2k-1,&\int_a^{\eta_i}\frac{\xi^{k-1/2}}{\sqrt{\xi-a}} \dd \xi= \frac{a^kC_{2k}^k}{4^k}\lsb -\ln a +2\ln\lb \sqrt{ \eta_i}+\sqrt{\eta_i-a}\rb + \sum_{j=1}^{k}\frac{4^j}{jC_{2j}^{j}}\lb \frac{\eta_i}{a}\rb^j\sqrt{1-\frac{a}{\eta_i}}\rsb. \label{eq:ttgenintmok}
\end{align}
\end{subequations}
They can be proven by using successive changes of variables $\xi=a+t^2$ and $t=\sqrt{a}\cot(s)$ and then use integral formula in Ref. \cite{bk:inttable}. Their correctness can also be directly checked numerically.

\end{widetext}

\end{document}